\documentclass{JHEP3} 
\usepackage{epsfig}
\epsfclipon

\newcommand{\nco}{\newcommand}
\nco{\beq}{\begin{equation}} \nco{\eeq}{\end{equation}}
\nco{\beqa}{\begin{eqnarray}} \nco{\eeqa}{\end{eqnarray}}
\nco{\lra}{\leftrightarrow}

\nco{\sss}{\scriptscriptstyle} \nco{\dphi}{\varphi}
\nco{\lsim}{\mbox{\raisebox{-.6ex}{~$\stackrel{<}{\sim}$~}}}
\nco{\gsim}{\mbox{\raisebox{-.6ex}{~$\stackrel{>}{\sim}$~}}}

\def\pref#1{(\ref{#1})}

\title{Asymmetrically warped compactifications\\ and gravitational Lorentz
violation}

\author{James M.\ Cline, Luis Valc\'arcel\\
Physics Department, McGill University,
3600 University Street, Montr\'eal, Qu\'ebec, Canada H3A 2T8\\
\email{jcline@physics.mcgill.ca},
\email{luis.valcarcel@mail.mcgill.ca}}

\preprint{McGill-03/28}


\abstract{
In a variant of the Randall-Sundrum braneworld model which has a charged black hole in an anti-de
Sitter bulk, the 4D speed of gravity depends upon one's location in the bulk, and in general differs
from the speed of light on a given 3-brane.  We apply phenomenological constraints on the difference in
the speeds of light and gravity to models which  use the warping to solve the hierarchy problem.  In
particular, consideration of the gravi-\v Cerenkov radiation of Kaluza-Klein gravitons by ultra-high
energy cosmic rays leads to a stringent limit on the fractional difference between the graviton and
photon speeds in RS-like models: it must be less than a part in $10^{23}$.}

\begin{document}

\section{Introduction}
The violation of Lorentz symmetry at high energies would be a revolutionary discovery, giving us important
clues about the nature of physics beyond the standard model \cite{Coleman}-\cite{Kostelecky}.  It has been
argued that such violations are predicted by some theories of quantum gravity \cite{Smolin}.  Although string
theory does not necessarily predict Lorentz violation, it can do so via background fields which lead to
noncommutative geometry \cite{Carroll}, or which explicitly violate Lorentz invariance in the gravitational
sector \cite{Frey}.  

Nevertheless, it is not easy to invent fundamental theories which break Lorentz symmetry in a plausible way.
Therefore it is useful to consider new possibilities for breaking Lorentz symmetry.  A novel approach was 
presented by Csaki {\it et al.} \cite{Csaki}, in the context of the Randall-Sundrum (RS) solution to the hierarchy
problem \cite{RS}, involving two branes separated by an extra dimension with the geometry of 5D anti-de Sitter
space.   Although the original RS model had Lorentz symmetry, ref.\ \cite{Csaki} (hereafter called CEG) considered a variant in which
the 5D bulk contains a black hole of charge $Q$ and mass $\mu$, giving a metric (the AdS Reissner-N\"ordstrom
black hole solution) with the line element
\beqa
\label{metric}
	ds^2 &=& -h(r)\, dt^2 + {r^2\over l^2} d\vec x^{\, 2} + {1\over h(r)} dr^2;\\
\label{h}
	h(r) &=& {r^2\over l^2} - {\mu\over r^2} + {Q^2\over r^4}.
\eeqa
The extra dimension has coordinate $r$, and the length scale $l$ characterizes the curvature of the AdS
geometry, which is determined by the negative bulk cosmological constant through $l^2 = -6/\Lambda$. (We work
in units where the 5D gravitational constant $\kappa_5=1$.)  

The presence of a bulk black hole (BH) can be physically motivated: it was argued in \cite{Hebecker} that emission of
gravitational radiation from a brane in the early universe can lead to the formation of a BH in the bulk. 
Possible cosmological consequences of gravity traveling faster than light in this scenario were considered
in references \cite{xxx}.
However in the cosmological solutions, the brane moves away from the BH as the universe expands, and as a
result the presence of  the BH quickly becomes irrelevant for the brane observer: its effects are redshifted
away. The presence of charge on the BH is motivated by the desire to have a solution in which the brane-BH
distance remains constant, and therefore its effects can also be important at late times.

Along any 4D slice of constant $r$, the metric \pref{metric} is Lorentz invariant, with speed of light
given by $dx/dt = (l/r) h^{1/2}$.  Thus an observer on a D-brane which is restricted to a particular value of
$r$ does not see direct evidence of Lorentz violation in the matter sector.  However gravity propagates in 
all dimensions, not just along the brane.  A 4D observer would find that gravity propagates with a different 
speed than matter.

It might be thought that the weakness of gravitational interactions would render this mild Lorentz violation
phenomenologically harmless; after all, the speed of gravity has not been measured.  However, there are  ways
in which it would be manifested.  One possibility is that self-energy diagrams with graviton loops would
induce speed differences between observable particles, like photons and electrons (where by ``speed" we mean
the maximum speed, as the momentum becomes infinite).  These kinds of constraints were considered in
\cite{Burgess}.

A more powerful constraint exists in the case where gravity travels slower than visible particles, especially
protons \cite{Moore}. The emission of gravitational \v Cerenkov radiation would very efficiently damp ultra
high-energy cosmic rays (UHECR's) to levels below those which are observed.  One of the main observations of
the present paper is that this is precisely the situation in asymmetrically warped models in which we are
assumed to be living on a negative tension brane, where the weak scale hierarchy can be solved by the
warping.  In CEG it was assumed that we live on a positive tension brane; in that the speed of gravity is
faster than that of light, so that the present strong constraint is not relevant.

In CEG, the asymmetric effects of warping were treated perturbatively.  In the present work we
also extend their analysis by considering these effects in the nonperturbative regime, and for the Kaluza-Klein
(KK) excitations of the graviton.  These could be relevant sources of Lorentz violation in colliders, since
in RS-like models the KK gravitons couple strongly to the standard model particles \cite{Rizzo}.

We begin by reviewing the CEG model and their perturbative derivation of the speed of gravity, and we
enumerate the most stringent experimental constraints on Lorentz violation in these models.
In section 3 we point out that the perturbative treatment can be misleading if the observer is located on the
positive tension brane, whereas it gives reliable predictions for observers on the negative tension brane.
In section 4 we show that the KK excitations of the graviton have similar Lorentz-violating properties to
the zero mode, considerably strengthening the \v Cerenkov bound introduced in section 2.

\section{The Models; perturbative speed of gravity}

\subsection{Model with two branes}

Similarly to the RS model, CEG cuts the bulk at two positions, call them $r_+$ and $r_-$, by the insertion
of branes and the use of $Z_2$ orbifold symmetry to define the discontinuity in the derivatives of the metric
functions at the branes.  The parts of the solution $r>r_+$ and $r<r_-$ are discarded, and replaced by 
mirror copies of the kept part of the solution to create a solution with orbifold fixed points at $r_\pm$.
Unlike RS, it is not possible to use conventional branes with equation of state
$p=-\rho$ at both positions.  There are four junction conditions which relate the energy densities $\rho_\pm$ and equations of 
state $p_\pm=w_\pm \rho_\pm$ to the black hole parameters: 
\beqa
\label{jc1}
	{\mu\,l^2} &=& 3\left( 1+ \frac{w_+}{36}{\rho_+^2 l^2} \right)r_+^4 = 
	3\left( 1+ \frac{w_-}{36}{\rho_-^2 l^2} \right)\!r_-^4 ;\\
	{Q^2\,l^2} &=& 2\left(\! 1\!+\! \frac{1\!+\!3w_+}{72}{\rho_+^2 l^2} \right)\!r_+^6 =  
	2\left(\! 1\!+\! \frac{1\!+\!3w_-}{72}{\rho_-^2 l^2} \right)\!r_-^6 .
\label{jc2}
\eeqa
It is easy to see that taking $w_+=w_-=-1$ implies that $r_+=r_-$.  Let us consider
the physically interesting case of a
large hierarchy, where 
\beq
	\epsilon \equiv r_-/r_+ \sim 10^{-16}.
\eeq
Physical masses on the negative tension brane
at $r_-$ (the ``TeV brane'') are suppressed relative to their bare values by the factor $\epsilon$, giving
a resolution of the weak scale hierarchy problem \cite{RS}.  By solving the jump conditions 
\pref{jc1}-\pref{jc2} for $\epsilon$, we see that there are two ways of achieving a small value of $\epsilon$:
either by (1) tuning the numerators $( 1+ \frac{w_+}{36}{\rho_+^2 l^2} )$, 
$(\! 1\!+\! \frac{1\!+\!3w_+}{72}{\rho_+^2 l^2})$
to be small, or (2) tuning the denominators  $( 1+ \frac{w_-}{36}{\rho_-^2 l^2} )$, 
$(\! 1\!+\! \frac{1\!+\!3w_-}{72}{\rho_-^2 l^2})$ to be large. 
\begin{enumerate}

\item
In the first case, $l\rho_-/6$ and $w_- (l\rho_-/6)^2$
can naturally be of order unity, while $w_+$ and $\rho_+$ must be tuned to the values
\beq
\label{wec1}
	w_+ \cong -1-2\epsilon^4\left( 1+ \frac{w_-}{36}{\rho_-^2 l^2} \right) < -1; \quad
	l\rho_+ = 6 + O(\epsilon^4),
\eeq
implying the relation 
\beq
\label{murel1}
	\mu = 	c_1{Q^2\over r_-^2}; \quad c_1 = { 3\left( 1+ \frac{w_-}{36}{\rho_-^2 l^2} \right)\over
	2\left(\! 1\!+\! \frac{1\!+\!3w_-}{72}{\rho_-^2 l^2} \right) } = O(1) .
\eeq 
The inequality in \pref{wec1} is a violation of the weak energy condition (albeit a small one), which
states that $p \ge -\rho$, and would thus require some exotic kind of stress energy on the Planck brane.

\item
In the second case, we need
\beq
\label{wec2}
	w_- = O(\epsilon^{2});\quad l\rho_- = O(\epsilon^{-6}) ,
\eeq
which implies
\beq
\label{murel2}
	\mu = 	c_2{Q^2\over r_+^2}; \quad c_2 = { 3\left( 1+ \frac{w_+}{36}{\rho_+^2 l^2} \right)\over
	2\left(\! 1\!+\! \frac{1\!+\!3w_+}{72}{\rho_+^2 l^2} \right) } = O(1) .
\eeq
\end{enumerate}

The relations \pref{murel1},\pref{murel2} are useful for computing the speed of electromagnetic radiation (or
any relativistic particles) on the branes.  Working
to first order in the Lorentz violating parameters $\mu$ and $Q^2$, the deviation in the speed relative to
unity is
\beq
\label{cem}
	\delta c_{\rm em} =  - {\mu\, l^2\over 2 r_\pm^4} + {Q^2 l^2 \over 2 r_\pm^6} =  {l^2\,Q^2\over 2 r_\pm^4}
	\left( {1\over r_\pm^2} - {c_i\over r_i^2}\right) ,
\eeq
where the choice $\pm$ refers to which brane one is on, and 
the label $i$ refers to the choice of tunings immediately above; hence $r_1 = r_-$ and $r_2 = r_+$.  

\subsection{Model with one brane and a horizon}

The preceding discussion assumed the existence of two branes bounding the extra dimensional space. It is also
possible to eliminate the TeV brane altogether by not cutting out the small $r$ region containing the black
hole.  In that case one should choose $\mu$ and $Q^2$ so that $h(r)$  vanishes at least once between $r=0$ and
$r=r_+$.  This insures there is a horizon shielding the black hole, in accordance with the cosmic censorship
hypothesis.  In this scenario we would necessarily be living on the Planck brane at $r=r_+$.  The parts
of the jump conditions \pref{jc1}-\pref{jc2} involving $\rho_+$ and $w_+$ would still be valid.  CEG showed
that it is not possible to embed the brane and still have a horizon shielding the black hole unless the 
weak energy condition is again violated, $w_+ < -1$.  This is most easily studied when the special relation
$4\mu^3 l^2 = 27 Q^4$ holds; this is the limiting case where the two horizons of the AdS Reissner-N\"ordstrom 
black hole degenerate into a single horizon, located at $r_h/l = ( 2Q/l^2)^{1/3}/\sqrt{2}$. 
One can solve the jump conditions to find that the brane's equation of state depends on its position,
$r_+$, and is given by 
\beq
	w_+ = -1  -2\left({r_h\over r_+}\right)^4 + O\left({r_h\over r_+}\right)^6 .
\eeq
Hence the violation of the weak energy condition becomes small as the brane is moved farther away from the
black hole. Ref.\ \cite{CF} showed that this violation could be moved into the bulk stress-energy tensor,
but never eliminated, so long as the horizon exists.  
The deviation in the speed of light on the brane in this scenario is given by
\beq
	\delta c_{\rm em}  = -\frac32 \left({r_h\over r_+}\right)^4 + \left({r_h\over r_+}\right)^6 .
\eeq
We mention this case for completeness, but we will not further consider how gravity propagates in this case.

\subsection{The speed of gravity; experimental constraint}

Now we turn to the speed of gravity.  Its determination is simplified by CEG's observation that the dynamics of
gravity are the same as those for a massless bulk scalar field, with action and equation of motion
\beqa
	S &=& \frac12\int d^{\,4}\!x\, dr \sqrt{-g} g^{AB}\partial_A\phi \partial_B\phi ;\\
\label{eom}
	{\delta S\over\delta\phi} &=& -\partial_A (\sqrt{-g} g^{AB}\partial_B\phi) = 0 .
\eeqa
The problem can be further simplified by splitting the metric into a pure AdS part and the perturbation
involving the black hole parameters.  In other words, one works perturbatively to first order in $\mu$ and
$Q^2$.  The solution for the $n$th KK mode is separable, $\phi_n(x^\mu,r) = 
e^{i(\omega_n t - \vec q\cdot \vec x)}\phi_n(r)$, where
$\phi_n(r)$ satisfies Neumann boundary conditions at the branes, since the jump conditions
\pref{jc1}-\pref{jc2} have already been satisfied by the background solution.
The dispersion relation of the graviton zero mode solution in the tower of KK excitations can be determined
analytically (in the model with two branes) for a solution with 3-momentum $\vec q$:
\beq
\label{disp} \omega_0 = |\vec q| + \delta\omega_0;\quad
	\delta\omega_0 = {|\vec q| \, l^2\over 4\, \epsilon^2}\left( -2{\mu\over r_+^4} + 
	{Q^2\over r_+^6} (1 + \epsilon^{-2}) \right) .
\eeq
The deviation in the speed of gravity from 1 is given by
\beq
\label{dcgrav}
	\delta c_{\rm grav} = {\partial\delta\omega_0\over \partial |\vec q|} .
\eeq

The relevant Lorentz-violating observable is the difference $\delta c_{\rm em} - \delta c_{\rm grav}$.
In general, the sign of this difference depends on the value of the $O(1)$
constant $c_1$ or $c_2$.  However for TeV brane observers, if the hierarchy is very large ($\epsilon\ll 1$),
and if we exlude the case \pref{wec1}-\pref{murel1} which violates the weak energy condition, 
the difference is completely dominated by the term $l^2 Q^2/(2r_-^6)$ in \pref{cem}, and leads to 
\beq
	\delta c_{\rm em}  - \delta c_{\rm grav}  = {l^2 Q^2\over 2r_-^6} .
\eeq
Because the difference is positive, the stringent constraints due to gravitational \v Cerenkov radiation of UHECR protons
is applicable \cite{Moore}:
\beq
	\delta c_{\rm em}  - \delta c_{\rm grav} < 2\times 10^{-15} \ (10^{-19}) .
\eeq
where the less stringent limit assumes the UHECR's are of galactic origin, and the more stringent one applies
if they originally come from neutrinos originating from cosmological distances.  Therefore we have the bound
\beq
\label{newbound}
{\delta h(r_-)\over h(r_-)} \cong {l^2 Q^2\over r_-^6} \lsim  4\times 10^{-15} \ (10^{-19}) .
\eeq
Notice that the quantity on the left is the fractional perturbation to the metric function $h(r)$ due to the
dominant Lorentz violating term, evaluated on the TeV brane: $h(r) \cong (r/l)^2 (1 + \delta h/h)$
It is therefore natural that this is the combination which is experimentally bounded.  \pref{newbound}
is one of the new constraints derived in this paper.

If we are willing to entertain the possibility of case 1, with exotic matter on the Planck brane which violates the
weak energy condition, then the difference of speeds as measured on the TeV brane becomes 
dependent on the details of the TeV brane stress energy,
\beq
	\delta c_{\rm em}  - \delta c_{\rm grav}  = {l^2 Q^2\over 2r_-^6}(1 - c_1) .
\eeq
If $c_1<1$, the \v Cerenkov bound again applies, in the obvious way.  If $c_1>1$, gravitational \v Cerenkov 
radiation cannot be produced.  Then the strongest bounds come from tests for deviations from general relativity
in the parametrized post-Newtonian (PPN) formalism.  A difference between the speed of light and that of
gravity implies the existence of a preferred reference frame, the one in which we have been working, where the
speeds of gravity and light are independent of direction.  In a boosted frame, one of these would become 
anisotropic, depending upon whether the boost respects Lorentz invariance of the gravitational or the
electromagnetic propagation.  (One but not both can be preserved, due to the difference in speeds.)
In such a situation, the parameter $\alpha_2$ of 
the PPN formalism is nonzero.  Essentially, we have two metrics, one for photons and one for gravitons,
so the effective theory is Rosen's bimetric one  \cite{Will}.  Such theories predict a torque which would cause precession
of the sun's spin axis.  The latter is closely aligned with the solar system's planetary angular momentum
vector.  If we assume that the alignment is not coincidental, then precession due to PPN effects is
constrained, and leads to the bound \cite{Nordtvedt}
\beq
\label{ssbound}
	|\alpha_2| = 2{|\delta c_{\rm em}  - \delta c_{\rm grav}|\over c_{\rm em}} < 1.2\times 10^{-7} ,
\eeq
which carries over to the quantity ${l^2 Q^2\over r_-^6}|1 - c_1|$.\footnote{We thank Guy Moore for
discussions on this point.}

\section{Nonperturbative analysis}
The above discussion assumed that the dispersion relation of the graviton gets modified in the simple way
which was predicted by treating the Lorentz-violating part of the metric, $\delta h$, to first order in
perturbation theory.  However, at large graviton momenta, the perturbative prediction can be modified.
To see this, one must examine the equation of motion \pref{eom}.  Let us rewrite it to first order in $\delta
h$, with $\phi_0^{(0)}$ and $\phi_0^{(1)}$ respectively denoting the zeroth and first order solutions,
in powers of $\delta h$, for the Kaluza-Klein zero mode.  We will also change coordinates to the form
$r/r_+ = e^{-ky}$ with $k=l^{-1}$.  The wave equation for the zero mode with 3-momentum $q$ is 
\begin{eqnarray}
	{\partial_y^2}{\phi_0^{(1)}}(y)-4k\, \mathop{\rm sgn}(y)\,\partial_y{\phi_0^{(1)}}(y) 
	&=& e^{2k|y|}\left(-2q\,\omega_0^{(1)}+{q^2}
	{\delta h\over h}(y)\right)\phi_0^{(0)}\\ &=& e^{2k|y|}\left(-2q\,\omega_0^{(1)}+{q^2\over k^2} \left(- \frac{{
	\mu}}{{ r_+^4 }}e^{4k|y|}+ \frac{{ Q^2}}{{ r_+^6 }}e^{6k|y|}\right)\right)\phi_0^{(0)}.
	\nonumber
\end{eqnarray}
Regardless of how small $\delta h/h$ is, at sufficiently large momenta the source term $q^2(\delta
h/h)\,\phi^{(0)}$ becomes so large that it can no longer be reliably treated as a perturbation.

To explore this, we have numerically solved the full, unperturbed graviton wave equation to find the
dispersion relation for the zero mode, $\omega_0(q)$, and compared this result to the perturbative 
prediction \pref{disp}.  To make the problem tractable, we have chosen definite values for the brane
stress energy parameters. For the tuning of parameters, we adopt case 2 above, eqs.\ \pref{wec2}-\pref{murel2}, 
so that $c_2$ depends on the positive tension brane parameters.  These we take to be 
 $w_+=-1$, while leaving  $\rho_+$ free to vary.  The full wave equation can be written as
\begin{equation}
\phi''(y) -4k\, \mathop{\rm sgn}(y)\,\left( \frac{1 -{\delta}\, e^{6k|y|}}{\hat h(y)} \right)\phi'(y)
+ e^{2k|y|}\left( \frac{\omega^2}{\hat h^2(y)}-\frac{q^2}{\hat h(y)} \right)\phi(y)=0,
\label{EOM}
\end{equation}
where 
\begin{equation}
\hat h(y)=1+ \delta\, e^{4k|y|} \left( -3+2e^{2k|y|} \right);\quad \delta \equiv  1 - 
{\rho_+^2 l^2\over 36} .
\end{equation}
In this form, it is clear that the Lorentz-violating effects, coming from $\hat h - 1$, are maximized at the
TeV brane, where $y = y_-$.  For the purposes of comparing exact results to the perturbative ones, we 
quantify the perturbation by the parameter
\beq
	\Delta \equiv 2\delta e^{6ky_-} .
\eeq
We expect the perturbative treatment to be valid when $\Delta \ll 1$.  

With these definitions we can now compare the exact numerical results with the perturbative ones. The
deviation of the  dispersion relation of the zero mode, relative to its Lorentz-conserving value,  is plotted
as a function of momentum for several values of $\Delta$, and as a function of $\Delta$ for several momenta,
in figure \ref{fig1}, where $\delta\omega$ and $q$ are given in units of $k = l^{-1}$.  We have taken a very
small hierarchy, $ky_- = 2$ for this illustration, and will comment on the extrapolation to larger values
below.  The interesting feature is that the true $\delta\omega(q)$ flattens to a constant value at large $q$,
contrary to the perturbative expectation.  

\FIGURE{
\epsfig{file=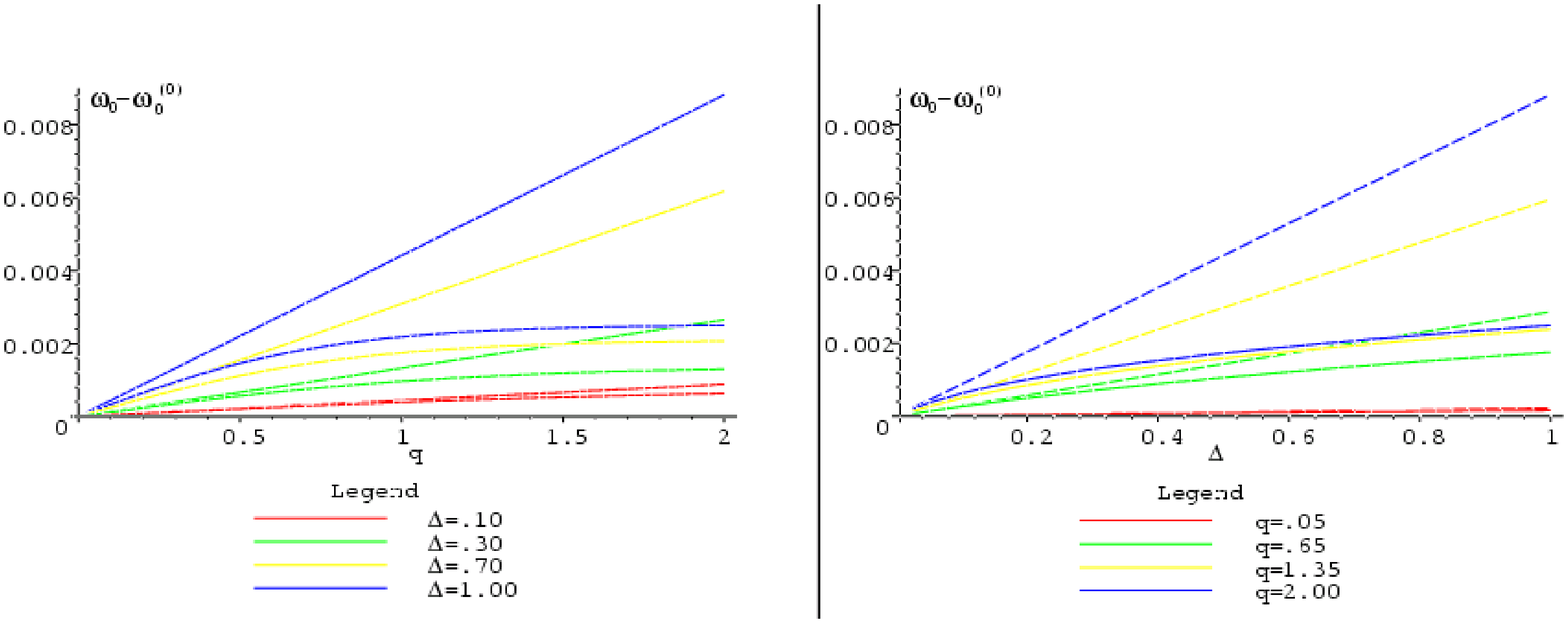,width=\hsize}
\caption[Correction to $\omega_0^{(0)}$ as a function of $q$ and $\Delta$]
{Correction to $\omega_0^{(0)}/k$ as a function of $q/k$ (left) and $\Delta$ (right).  The
upper (straight) lines show the perturbative prediction (\ref{disp}), while the lower (curved) lines show the numerical
result.  Each color represents a different perturbation or momentum, and $y_-=2/k$.}
\label{fig1}}

If we compare the speed of gravity to that of hypothetical photons which are trapped on the {\it positive
tension brane}, a consequence of this behavior of the graviton dispersion relation is that the speed
difference  (with speed defined using the group velocity $v = d\omega/dq$)  does not remain constant at large
momenta, but rather vanishes as $q\to\infty$, as shown in figure \ref{fig2}.  At large momenta, gravitons tend
toward the same speed as radiation on the Planck brane.  This can be understood as follows.  The graviton zero
mode is localized on the Planck brane.  At large $q^2$, the effect of the terms $(  \frac{\omega^2}{\hat
h^2(y)}-\frac{q^2}{\hat h(y)})$ in the equation of motion \pref{EOM} is to localize it even more, thus driving
the graviton to resemble more closely radiation which is trapped on the Planck brane.  This trend becomes
evident for momenta $q\gsim 1/l$, where the speed difference is significantly reduced relative to its maximum
value at $q=0$.  

As the hierarchy between the Planck brane and the negative tension brane is increased by letting $y_-$ go to 
larger values, the Lorentz violating effects seen by a Planck brane observer are suppressed, if the parameter
$\Delta$ is held fixed.  The magnitude of $c_{\rm grav} - c_{\rm em}$ scales like $e^{-2ky_-}$.  This
can be understood analytically, using the perturbative results of the previous section.  One can show that,
for the present choice of parameters, 
\beq
	c_{\rm grav} - c_{\rm em} \ =\  \frac12\left(\cosh(2ky_-)-1\right)e^{-4ky_-} \Delta \ 
 \sim\  +O(e^{-2ky_-}) .
\eeq
At the same time, the $q$-axis in figure \ref{fig2} gets rescaled by a factor of $e^{-ky_-}$, such that
the range of momenta with an appreciable deviation in $c_{\rm grav} - c_{\rm em}$ gets shifted to smaller
physical values.  In the hierarchy-solving case $e^{-ky_-}\sim 10^{16}$, this corresponds to the TeV scale.

\FIGURE{
\epsfig{file=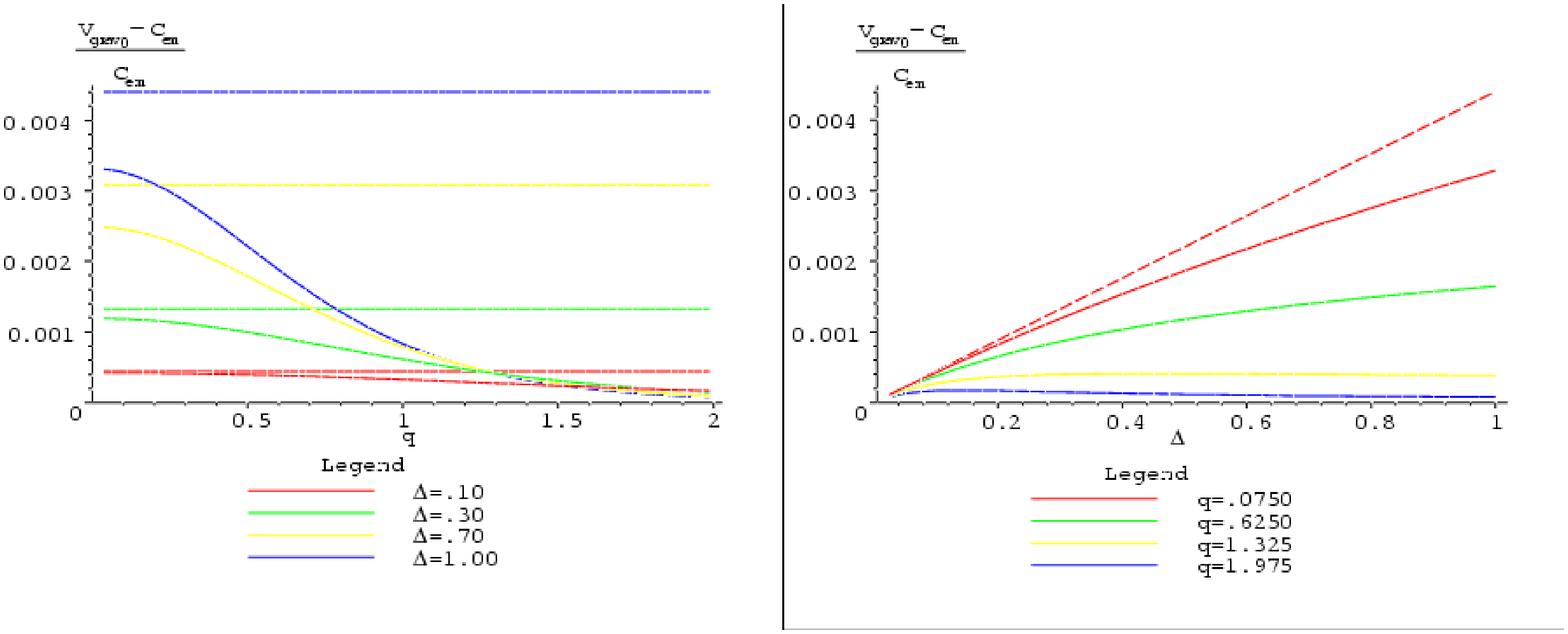,width=\hsize}
\caption[Relative speed difference between the gravity zero mode and the photon as a function of $q$ and $\Delta$ when the physical brane has positive tension] 
{Relative speed difference between the gravity zero
mode and a photon on the {\it positive tension brane} as a function of $q$ (left) and $\Delta$ (right) 
when the physical brane has positive tension.  The
upper (straight) lines show the perturbative prediction derived from (\ref{dcgrav}), 
while the lower (curved) lines
show the numerical result.  Each color represents a different perturbation or momentum, and $y_-=2/k$.
In the right panel, all the perturbative curves coincide with each other.}
\label{fig2}}

On the other hand, an observer on the TeV brane will not be concerned with this small momentum-dependence in
the graviton speed, because the difference between $c_{\rm grav}$ and $c_{\rm em}$ remains relatively  large
even at high momenta, as shown in figure \ref{fig3}.  Moreover, in contrast to the case of the Planck brane
observer, the difference $c_{\rm grav}-c_{\rm em}$ remains constant as the hierarchy is increased, if
$\Delta$ is held fixed.  Again, this can be understood by computing the speed difference perturbatively,
for the present choice of parameters,
\beq
	c_{\rm grav} - c_{\rm em} \ =\   -(\cosh(2ky_-)-1)e^{-2ky_-} \Delta \ \sim\  
	-O(1)\times\Delta .
\eeq

\FIGURE{
\epsfig{file=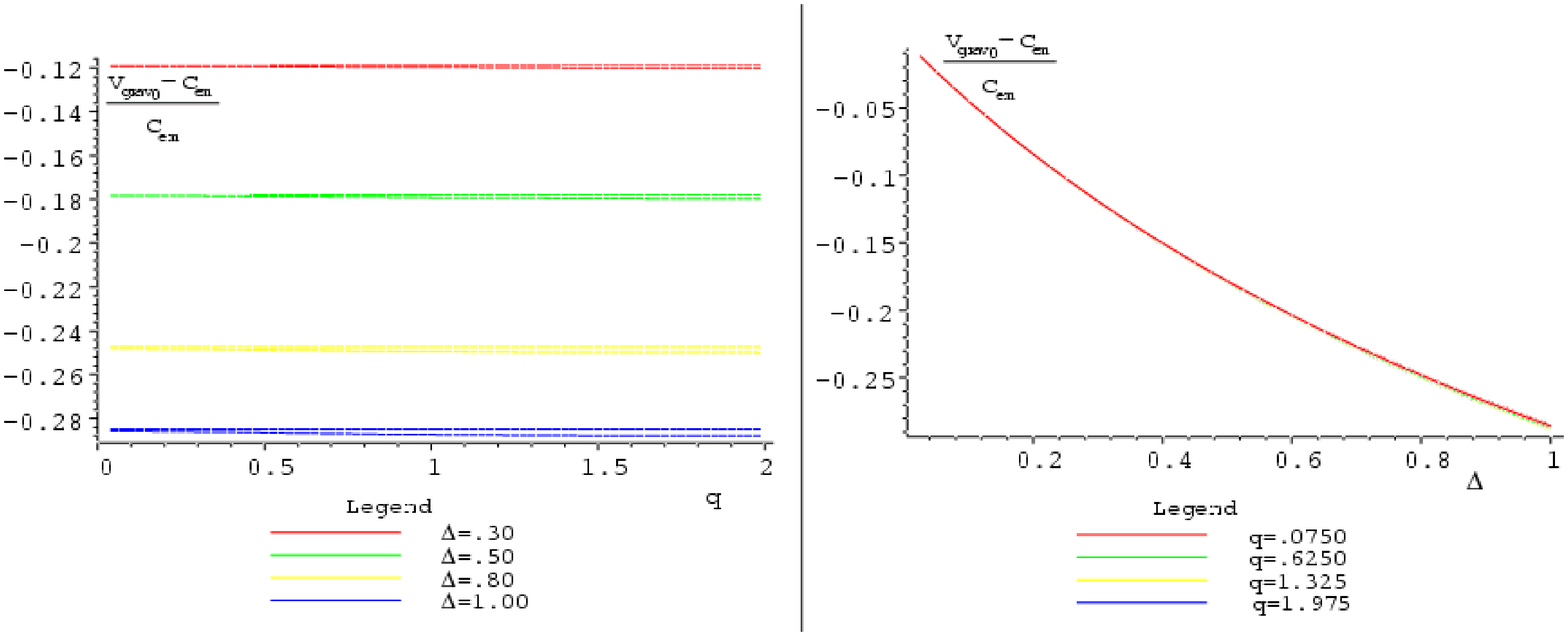,width=\hsize}
\caption[Relative speed difference between the gravity zero mode and the photon as a function of $q$ and $\Delta$ when the physical brane has negative tension]
{Same as figure 2, but here the photon is on the negative tension (TeV) brane.  All curves are essentially
coincident in the right panel.} 
\label{fig3}}

The conclusion of this section's analysis is that we can trust the results of the perturbative treatment if we
assume that observers are living on the negative tension brane, as one would expect if the hierarchy problem is being
addressed.  Only for Planck brane observers would it be important to distinguish the perturbative from the
exact results.

\section{Lorentz-violating KK modes}

It would be interesting if Lorentz violating kinematics could be observed in the laboratory, in particle
collider experiments.  In section 2 we noted that the speed difference between gravity and light on the  TeV
brane should be less than a few parts in $10^7$.  Such a small effect would probably be impossible to see at 
the LHC, even though KK gravitons interact strongly (with TeV-suppressed instead of Planck-suppressed 
couplings) with standard model particles and would be copiously produced, if sufficiently light.

Even if their Lorentz-violating properties cannot be directly detected, an interesting 
indirect effect is the \v Cerenkov emission of KK gravitons from UHECR's, which would strengthen
the bound mentioned in section 2.  That bound conservatively counted only the damping due to emission of
massless gravitons.  But KK modes can also be emitted, so long as their mass obeys the inequality
\beq
\label{Cer}
	{m^2_{\rm KK}\over 2\vec p^{\,2} }< c_{\rm em} - c_{\rm grav}(p) ,
\eeq
where $\vec p$ is the momentum of the UHECR.  This is the condition for \v Cerenkov emission to be
kinematically allowed. In the situation where the warp factor
$\epsilon = 10^{-16}$, the KK mass gap is of order TeV.  The highest energy cosmic ray which has been
detected had energy $3\times 10^{11}$ GeV \cite{Bird}; thus for a speed difference that saturates the
solar system bound \pref{ssbound}, the number of relevant KK modes is of order $10^5$.  It is therefore worth
exploring whether the KK modes have similar Lorentz-violating kinematics as the graviton zero mode.

We have addressed this problem both numerically and analytically.  The analytic approach is to solve the
wave equation once again treating $\delta h$ to first order in perturbation theory, but now expanding around
the zeroth order Lorentz-conserving KK wave function, $\phi_n(y) \cong \phi_n^{(0)}(y) + \phi_n^{(1)}(y)$,
in order to find the corresponding energy eigenvalue, $\omega_n(q) \cong \omega_n^{(0)}(q) + 
\omega_n^{(1)}(q)$.  The solution is described in detail in the Appendix.  We have also numerically solved
the full wave equation using the shooting method, to find the dispersion relation $\omega_n(q)$.  The
deviation of $\omega_n(q)$ from the standard result $\omega=q$ is shown for the first two KK modes in
figure \ref{fig4}.  The comparison between the perturbative and numerical results is shown there.

\FIGURE{
\epsfig{file=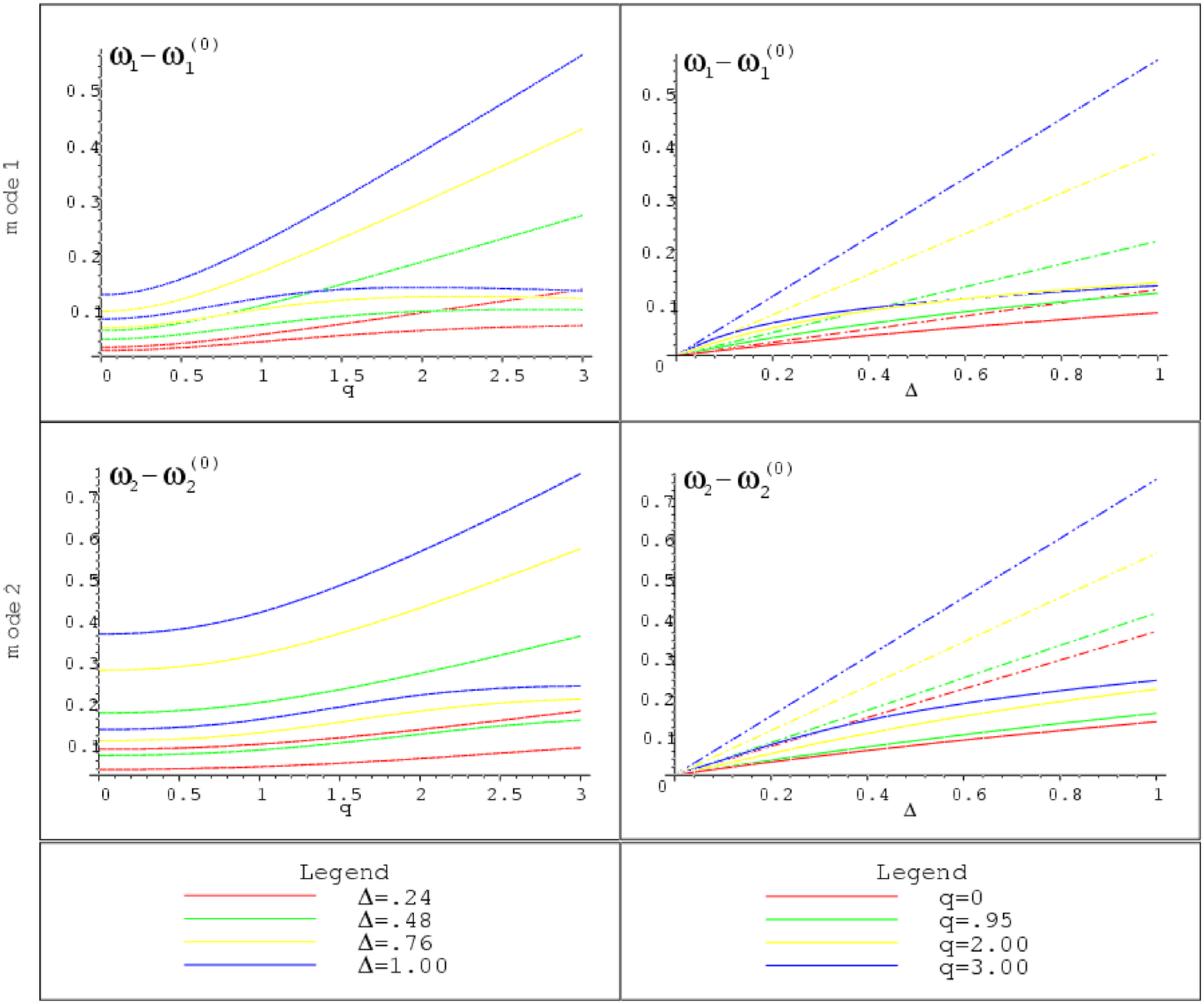,width=\hsize}
\caption[Correction to $\omega_n^{(0)}$ as a function of $q$ and $\Delta$]
{Same as figure 1, but for the first two massive KK modes.  The perturbative prediction comes from
eq.\ (\ref{eq:omega1}).}
\label{fig4}}

As we did for the KK zero mode, we can compute the difference in the speed of gravity relative to that of
particles trapped on the positive tension brane.  In order to highlight the differences which are
due to Lorentz violation, we take the conventional trapped particles to have the same rest mass as that of 
the KK graviton to which it is being compared.  The results, shown in figure \ref{fig5}
are qualitatively similar to those for the zero mode: gravity is faster than particles on the Planck brane,
but the speed difference becomes smaller at higher momenta.

\FIGURE{
\epsfig{file=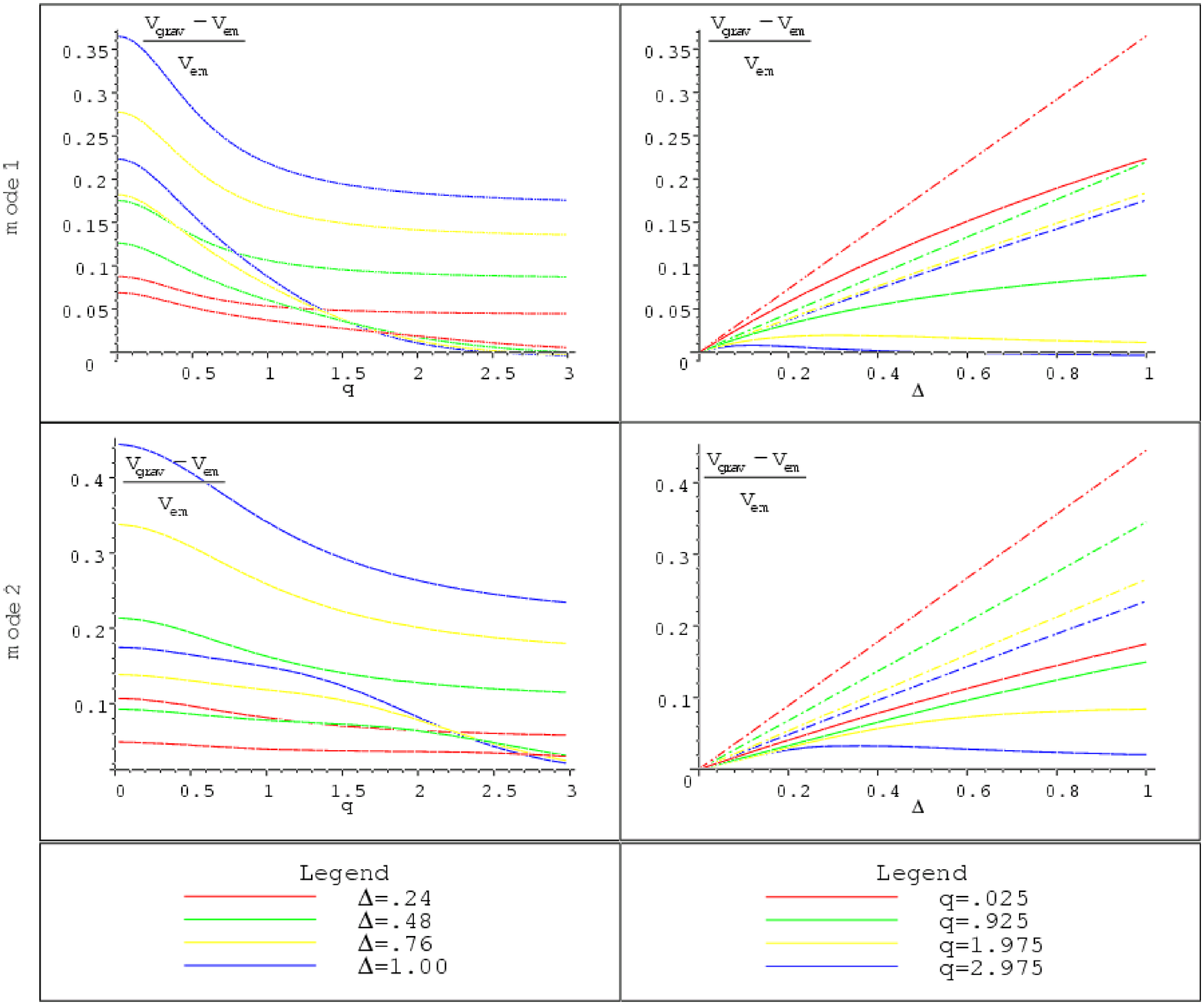,width=\hsize}
\caption[Relative speed difference between the gravity KK modes and hypothetical particles of the same masses,
living on the positive tension brane, 
 as a function of $q$ (left) and $\Delta$ (right).]
{Relative speed difference between the first two massive gravity KK modes and hypothetical particles of the same masses,
living on the positive tension brane, 
 as a function of $q$ (left) and $\Delta$ (right). Upper curves are perturbative, lower are numerical
results.}
\label{fig5}}

\FIGURE{
\epsfig{file=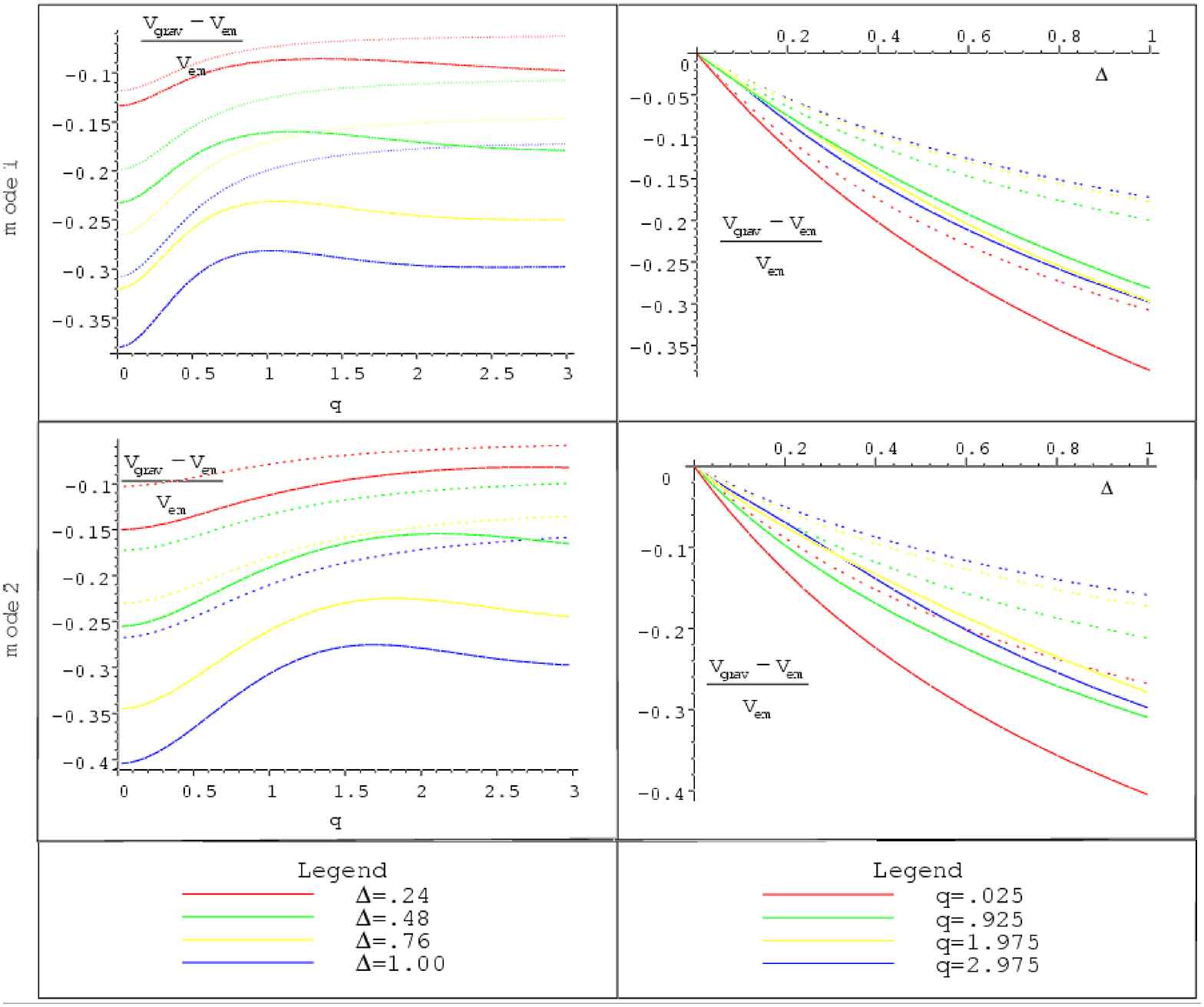,width=\hsize}
\caption[Relative speed difference between the gravity KK modes and hypothetical particles of the same masses,
living on the negative tension brane, 
 as a function of $q$ (left) and $\Delta$ (right).]
{Relative speed difference between the gravity KK modes and hypothetical particles of the same masses,
living on the negative tension brane, 
 as a function of $q$ (left) and $\Delta$ (right). Upper curves are perturbative result, lower are numerical.}
\label{fig6}}

More interesting is the speed difference relative to particles on the TeV brane.  Figure \ref{fig6} shows that,
again like the zero mode, KK gravitons are slower than same-mass particles on the TeV brane.  Therefore
our expectation that they are produced in gravi-\v Cerenkov radiation by particles with momenta satisfying
\pref{Cer} is justified, and we should revise the bound.  Adapting the result of \cite{Moore} to the tower
of TeV-mass-gap gravitons whose couplings are only TeV-suppressed, the rate of energy loss of a UHECR
with momentum $p$ and mass $m_p$ is given by
\beq
	{dE\over dt} \sim {p^2\over {\rm TeV}^3} \int_0^{q_{\rm max}} dq\, q \int_0^{M_{\rm max}} dM\, 
	\sin^4\!\theta ,
\eeq
where $q$ is the momentum of the emitted KK graviton, $M$ is its mass, and $\theta$ is the angle of emission
of the \v Cerenkov radiation.  The kinematic limits and $\theta$ (in the limit $\theta\ll 1$) are given by
\beqa
	\theta^2 &=& (1-q/p)q^{-2}(M_{\rm max}^2 - M^2)\\
	M^2_{\rm max} &=& q^2\left( 2\,\delta c - {m_p^2\over p^2(1-q/p)} \right)\\
	q_{\rm max} &=& p\left(1 - {m_p^2\over 2 p^2\,\delta c}\right) .
\eeqa
Here $\delta c = c_{\rm em} - c_{\rm grav}$, which must be positive in order for \v Cerenkov emission to occur.
Evaluating the integrals gives
\beq
	{dE\over dt} \sim 0.1\, (\delta c)^{5/2} f(\delta c)\,{p^5\over {\rm TeV}^3} ,
\eeq
where the function $f$ is approximately 1 for $\delta c\gg {m_p^2\over 2p^2}$ and drops very rapidly to
zero (like $(1-m_p^2/2 p^2\delta c )^{11/2}$)
as $\delta c\to {m_p^2\over 2p^2}$.  The bound on $\delta c$ comes from demanding that $dE/dt$ be less 
than $p/L$ for a cosmic ray which is propagating over a distance $L$.  For the highest energy cosmic ray 
observed,
which is identified with a proton, the bound is satisfied by taking $\delta c$ to be below the kinematic
limit where $f=0$.  This implies that
\beq
	\delta c < {m_p^2\over 2p^2} \cong 10^{-23} .
\eeq

\section{Conclusions}
Asymmetric warping  can provide a plausible means of introducing Lorentz violation into a theory with extra
dimensions, which is essentially a form of spontaneous breaking due to the gravitational background.  Since
at tree level the violation is confined to the gravitational sector, the effects can be sufficiently weak
to be at the borderline of detection.  In this paper we have explored some of the consequences of a theory
where the asymmetric warping comes from the presence of a charged black hole in the extra dimension.  Such
a scenario can be compatible with the RS solution to the weak scale hierarchy problem.  Unlike RS, in the
black hole case it is necessary to allow for nonstandard equations of state for the tensions of at least one
of the branes, though it is still possible to respect the weak energy condition.  One issue we have not
explored is the role of a stabilization mechanism such as that of Goldberger and Wise \cite{GW} for the
size of the extra dimension.  One might expect that the bulk is already stabilized once the brane energies
and equations of state are fixed, since algebraically the ratio of brane positions $r_-/r_+$ is determined.
However it is possible the radion mass$^2$ is negative, even though it is not zero.  Even if it has
the right sign, the distortions of the bulk geometry by a bulk scalar could have an effect on the propagation
of gravity.  It would also be interesting to know whether pure tension branes could be admitted with the
addition of a bulk scalar, as it would be desirable to eliminate the need for an unusual equation of state.
Finally, we did not consider how Lorentz violation is manifested in the single brane case,
mentioned in section 2.2, although we expect it to be similar to the case of  a Planck-brane observer
in the two brane model.  These subjects could merit further study.

\acknowledgments
We thank Guy Moore for very helpful discussions.

\appendix

\section{Perturbative KK mode solution}
The zeroth order in Lorentz-violation KK graviton wave function 
(which like the zero mode obeys Neumann boundary conditions) is \cite{GW0}
\begin{equation} 
	\phi_n^{(0)}(y)=\frac{e^{2k|y|}}{N_n} \left( J_2 \left(\frac{m_n e^{k|y|}}{k} \right) -
	\frac{J_1\left(\frac{m_n}{k} \right)}{Y_1 \left(\frac{m_n}{k} \right)} Y_2 \left(\frac{m_n
	e^{k|y|}}{k} \right)  \right) ,
\end{equation}
where $N_n$ is a normalization constant.
The $m_i$ are determined by the boundary conditions; in the limit of a large hierarchy, they satisfy
$J_1(m_n e^{ky_-}/k) = 0$.  

The wave function correction $\phi_n^{(1)}(y)$ is difficult to compute, but as is familiar from perturbation
theory in quantum mechanics, we don't really need it if we are only interested in the correction to the
eigenvalue, $\omega_n^{(1)}(q)$.  In the quantum mechanical analogy, we require only the matrix element
of the perturbation to the Hamiltonian.  Defining $\psi_n(y)=e^{-2k|y|}\phi_n^{}(y)$, the equation of 
motion becomes
\begin{equation}
	H\psi_n(y)=-\frac{\omega_n^2e^{2k|y|}}{\hat h(y)}\psi_n(y) ,
	\label{HermitianEOM}
\end{equation}
where, to first order in the perturbation,
\begin{eqnarray}
	H &=& \partial_y(\hat h(y)\partial_y)-4k^2\hat h(y)+2k\mathop{\rm sgn}(y)\,\hat h'(y)-q^2e^{2k|y|}+4k\hat h(y)\delta(y)\\
	&=&\left[\partial_y^2 -4k^2  -q^2e^{2k|y|}+4k \delta(y)\right] \nonumber\\
	& &+ \left[\partial_y(\hat h^{(1)}(y)\partial_y)- \left(4k^2 - 4k\delta(y)\right)
	\hat h^{(1)}(y) +2k\mathop{\rm sgn}(y)\,\hat h'^{(1)}(y)\right] . \\
	&\equiv& H^{(0)}+ H^{(1)},
\end{eqnarray}
with $\hat h^{(1)} =\hat h - 1$.

We take the inner product with $\psi_n^{(0)}$ and use the Hermicity of $H^{(0)}$ to cancel the terms involving $\psi_n^{(1)}$.  This yields
\begin{equation}
	\omega_n^{(1)}=\frac{-\langle\psi_n^{(0)}|H^{(1)}| \psi_n^{(0)}\rangle + 
	(\omega_n^{(0)})^2\langle\psi_n^{(0)}|{\hat h^{(1)}}e^{2k|y|}| 
	\psi_n^{(0)}\rangle}{2\omega_n^{(0)}\langle\psi_n^{(0)}|{e^{2k|y|}}| 
	\psi_n^{(0)}\rangle}.
\end{equation}
The inner product in the denominator is unity, with the appropriately normalized wave function.
Putting in the explicit form of $H^{(1)}$, the expression becomes
\begin{eqnarray}
	\omega_n^{(1)}&=&-\frac{1}{\omega_n^{(0)}}\int_0^{y_-}\psi_n^{*(0)}\left(\partial_y(\hat h^{(1)}
	\partial_y)-4k^2\hat h^{(1)}+2k\hat h'^{(1)}\right) \psi_n^{(0)}\,dy \nonumber \\
	& & + \omega_n^{(0)}\int_0^{y_-}{\hat h^{(1)}} e^{2ky} |\psi_n^{(0)}|^2\,dy \\
	&\equiv&-\frac{1}{\omega_n^{(0)}}A_n+ \omega_n^{(0)}B_n.
	\label{eq:omega1}
\end{eqnarray}
Since $\omega_n^{(0)}=\sqrt{m_n^2+q^2}$, and $m_n/k\propto e^{-ky_-}$,  all the terms are roughly comparable for $q \ll m_n$.  But, for large momentum $q$, only the second integral is important.  In that case,
\begin{eqnarray}
	\omega_n^{(1)}&\approx& q\int_0^{y_-}|\psi_n^{(0)}|^2 {\hat h^{(1)}}e^{2ky} \,dy \\
	&=& {q\over k^2}\int_+^{y_-}|\psi_n^{(0)}|^2 \left(- \frac{{ \mu}}{{ r_+^4 }}e^{6ky}+ \frac{{ Q^2}}{{ r_+^6 }}e^{8ky} \right) \,dy \\
	&\cong& \frac{q}{k^2 N_n^2}\int_+^{y_-} J_2^2 \left(\frac{m_n e^{ky}}{k} \right) \left(- \frac{{ \mu}}{{ r_+^4 }}e^{6ky}+ \frac{{ Q^2}}{{ r_+^6 }}e^{8ky} \right) \,dy 
\end{eqnarray}
where the last approximation holds in the limit of a large hierarchy.
The resulting integral can only be done numerically. However, our
analysis is still useful since it tells us that the first correction term to $\omega_n$ is
proportional to $q$ for large momentum, just as for the zero mode.  Moreover we can show
how the dispersion relation is expected to change.  Squaring $\omega_n$ gives 
\begin{equation}
	\omega_n^2=(\omega_n^{(0)}+\omega_n^{(1)}+\cdots)^2 \cong (m_n^2+q^2)(1+2B_n-2A_n)
\end{equation}
to first order in the perturbation.  Comparing with the usual dispersion relation, one sees that the limiting speed 
of the $n{\rm{th}}$ mode has been modified to 
\begin{equation}
  c_{\rm{grav}_n}^2=1+2B_n ,
 \label{eq:cgravPert}
\end{equation}
and that the mass of the mode changes due to the presence of the black hole perturbation:
\begin{equation}
	M_n^2=m_n^2(1+2B_n)-2A_n.
	\label{eq:FirstOrderMass}
\end{equation}
In terms of these, the group velocity of the $n{\rm{th}}$ mode with momentum $q$ is
\begin{equation}
 v_{\rm{grav}_n}  = \frac{\partial \omega_n}{\partial q} \cong 
 \frac{c_{\rm{grav}_n}^2 q}{\sqrt{M_n^2+c_{\rm{grav}_n}^2 q^2}}. \label{eq:VgravPert}
\end{equation}

As a consistency check, it can be verified that this result reduces to the previous one for the zero mode case
where it is possible to do the integrals.

\end{document}